# Reaching for equilibrium : an alternative view of the unbalanced carbon cycle.

By François Ouellette

## Abstract

A model is proposed to explain the observed correlation between monthly fluctuations in atmospheric $CO_2$ concentrations and temperatures. The model relies on the oceans being in a temperature-dependent equilibrium with the atmosphere. When temperature changes, the system attempts to restore the equilibrium, with a time constant constrained by the dynamic fluxes between the two. The model is used to reconstruct both the historic evolution of $CO_2$ concentrations since 1885, as well as the monthly fluctuations since 1959. The best fit is obtained with the southern hemisphere sea surface temperature data. In this new picture, human emissions play no role in the build-up of atmospheric $CO_2$, but do contribute to the fluxes that restore the equilibrium.

## Introduction

The two cornerstones supporting the Anthropogenic Global Warming (AGW) theory are undoubtebly the twin observations of a global temperature increase since the early 20th century, and the gradual rise in atmospheric $CO_2$ concentration since the beginning of the industrial age, as indicated by proxies, and by actual measurements taken at the Mauna Loa station and other stations since 1958. $CO_2$ being a potent greenhouse gas, it follows logically that its increasing concentration can explain the concurrent warming. Estimations of $CO_2$ forcing used in Global Circulation Models, together with other assumptions about, for example, water vapor feedback, and aerosol forcing, all concur in making a convincing case that human emissions are responsible for the warming.

However, that theory also rests on another taken for granted assumption: that the human emissions of $CO_2$ are themselves responsible for the build-up of $CO_2$ in the atmosphere. Indeed there seems to be a correlation between emissions and $CO_2$ concentration, and the isotopic carbon ratio indicates that about half of the emissions are still stored in the atmosphere, with the rest having been mostly uptaken by the oceans (Sabine et al. 2004). However, some intriguing questions about the carbon cycle have to this day remained unanswered:

- From the best estimates of human emissions, it appears that the atmospheric $CO_2$ concentration does not increase by the same amount annually, but rather by a fraction of it, the so-called "airborne fraction", which is about 55%. The mechanism by which the other constituents of the carbon cycle manage to absorb the rest of the emissions, and neither more nor less, is not known.

- $CO_2$ concentration does not rise synchronously with human emissions. While human emissions are a monotonically increasing function of time, the rate of $CO_2$ concentration change varies wildly on a monthly basis, by as much as the total human emissions.

It should also be noted that the carbon cycle is highly dynamic. Human emissions are but a small fraction (about 3%) of the total exchanges between the atmosphere, the land biomass, and the oceans. In effect, the human perturbation is small. Yet, it seems that it is sufficient to derail the cycle that would otherwise be in perfect equilibrium, as it has been for millenia.

To resolve the mystery of the monthly fluctuations in $CO_2$ uptake, many efforts have been made to try to quantify in detail the various fluxes between the carbon cycle constituents, and to model the movement of $CO_2$ in the atmosphere (Bousquet et al. 2000). In particular, it is recognized that the airborne fraction is influenced by major climatic events (Denman et al., 2007). Indeed, it has been recently pointed out (MacRae2008) that the monthly change in atmospheric concentration is rather well correlated with the global temperature anomaly. One simple way of showing this is to plot the temperature anomaly versus the change in $CO_2$, make a linear regression, and re-plot those two variables as a function of time, using the regression parameters to re-scale one of them. An example is shown on Fig. 1. For that purpose, and for the rest of this study, I have used, for the monthly rate of change in teh $CO_2$ concentration,, C, an annualized value $\Delta C$ defined as:

$$\Delta C (t) = C (t + 6 \text{ months}) - C (t - 6 \text{ months}) \qquad (1)$$

That definition avoids the effect of seasonal fluctuations. Of course, it has the drawback of not being the actual instantaneous monthly fluctuation. The only alternative would be to model those seasonal fluctuations, as well as the instantaneous trend, and numerically remove them, but this would also be prone to errors. The definition of Eq.(1), by virtue of the Mean Value theorem, assures that $\Delta C$ represents the true derivative somewhere in the 12 month interval. Because of that, the actual amplitudes in the fluctuations of $\Delta C$ should be relatively well captured.

The data for the global temperature anomaly were taken from the Hadley Center (http://www.cru.uea.ac.uk/cru/data/temperature/), and the data for $CO_2$ were from the Mauna Loa station, found at ftp://ftp.cmdl.noaa.gov/ccg/co2/trends/co2_mm_mlo.txt . From Fig.1, it can be seen that the $\Delta C$ curve reproduces surprisingly well the main features of the temperature anomaly curve. On the other hand, the two curves tend to depart at more recent dates, as if the $\Delta C$ curve lacked the trend present in the anomaly curve.

This exercise assumed a simple, linear correlation model where :

$$\Delta C = \alpha \, \Delta T + \beta \qquad (2)$$

However, this is not the only way to model the relationship between the two variables, that would lead to a similar behavior. Although there have been other attempts at

p.2

modeling the carbon cycle to emphasize the role of temperature on the various fluxes (Rörsch 2005), I will present, in this brief communication, a model that has the advantage of being extremely simple, and having an extremely good fit with both the observations of historical $CO_2$ concentrations, and monthly fluctuations in said concentration. The model has, however, surprising consequences about the possible cause of $CO_2$ accumulation in the atmosphere.

## Model

The carbon cycle is the result of a dynamic exchange of $CO_2$ between three major components: the oceans, the atmosphere, and the biomass. Of those three, the oceans are an immense sink of $CO_2$, holding as much as 38,000 PgC, as compared with 760 PgC in the atmosphere, and 2000 PgC in the land biomass. The annual exchange between the surface ocean layer and the atmosphere is about 100 PgC in both directions, and that between the atmosphere and the land biomass is 120PgC. Human emissions are only about 6.5 PgC per year. Both observations (Sabine et al. 2004) and models (Sarmiento et al. 1998) point to a predominant role of the oceans in long term carbon uptake. The uptake of carbon in the oceans is mostly through the "solubility pump", where carbon is dissolved in the cold ocean waters at high latitudes, and outgassed from the warm waters at low lattitudes (Raven et al. 1999).

Given the large ocean sink, one could portray the oceans as a big electrical capacitor, in equilibrium with a smaller capacitor, the atmosphere. The steady state atmospheric concentration is the result of that equilibrium. Over long time scales, the ocean sink is large enough to absorb or emit carbon dioxide at will to maintain the equilibrium. This equilibrium is known to be both temperature- and salinity-dependent, since $CO_2$ solubility is itself temperature dependent (Raven et al. 1999, Denman et al. 2007, Sarmiento et al. 1998). Now, because the exchanges between the two consituents are very dynamic, any perturbation to that equilibrium should immediately be reflected in the fluxes, just like a change in the voltage of a capacitor immediately induces a current.

The electrical analogy is in effect that of two charged capacitors, connected by a resistor, with one of the capacitances being temperature dependent. As that capacitance varies, it induces an imbalance, and an electrical tension between the two. Charges from the other capacitor will move to restore the balance of charges, inducing a current that will last until the tension has dropped to zero. In the case of the oceans and the atmosphere, the net fluxes between the two are perturbed until equilibrium is reached again. Therefore this simple model does not attempt to determine the temperature dependence of the various fluxes themselves, but rather assumes that the temperature-dependent equilibrium concentration of $CO_2$ in the atmosphere is regulated by the fluxes between the atmosphere and the oceans. The magnitude of the fluxes is what ultimately limits the response time of the system, much like the resistor value limits the rate of a capacitor discharge. Mathematically, the change in $CO_2$ concentration can therefore be described by those two simple equations:

$$\Delta C = (C_e(\Delta T) - C) / \alpha \tag{3}$$

$$C_e(\Delta T) = C_{e0}(1 + \gamma \Delta T) \tag{4}$$

where $\Delta C$ is the change in $CO_2$ concentration, $C_e(\Delta T)$ is the equilibrium concentration of $CO_2$ in the atmosphere at a temperature anomaly $\Delta T$, which is given as a linear departure from a "base" concentration $C_{e0}$ for a null temperature anomaly, at a rate $\gamma$. The sensitivity $\gamma$ is expressed as a fractional change in concentration per $^oC$. The time constant $\alpha$ is expressed in years. Thus the model has only 3 free parameters: $\alpha$, $\gamma$, and $C_{e0}$.

The fact that the system responds immediately does not mean, however, that equilibrium is reached immediately. Indeed the time constant can be very long, just like that of a very large capacitor. The time constant is dictated by the largest sink, which would presumably be the oceans, since they hold as much as 60 times the amount of carbon in the atmosphere, and by the magnitude of the annual fluxes between them and the atmosphere. I will assume here that the time constant is of the order of several decades, as it appears to be the sort of time scale required for carbon to reach the deeper ocean depths (Sabine et al. 2004).

With that simple model in hand, there are two things that can be done with it. First, the monthly change in atmospheric $CO_2$ concentration can be compared with that expected from the model, using the monthly temperature anomaly, over the 1959-2007 period. But assuming that the behavior described by eqs. (3) and (4) is a fundamental and permanent property of the system, one can also "reconstruct" the history of $CO_2$ concentration in the atmosphere, using the historical temperature data. Since there are relatively reliable global temperature data since about 1885, the reconstruction is not limited to the period 1959-2007. However, it should be made to fit the $CO_2$ measurements over the latter period. To perform such a reconstruction, I used the mean annual temperature anomaly to calculate the annual change in concentration with eqs. (3) and (4), and add it to the year's concentration to determine that of the next year, and so on for the entire period.

One can already foresee a possible consequence of that model : if the global temperatures increase over a long period, at a rate faster than the response time, the system will be out of equilibrium over the whole period, and the $CO_2$ concentration will steadily increase. Since, in effect, the Earth has been warming ever since the early 20$^{th}$ century, the model would predict such a gradual rise in $CO_2$ concentration. But that rise is the **result** of the warming, and not its cause. But let us see first if the model does fit the observations or not.

## Results

For the simulations, the initial $CO_2$ concentration in 1885 was set arbitrarily at 280 ppmv. That value is deemed by most to be the "pre-industrial" $CO_2$ concentration. Although the Hadley Centre temperature data go back as far as 1850, the uncertainty prior to 1885 is probably too large to warrant a reliable use. Since the model requires the integration of all the annual carbon dioxide changes, the errors would inevitably accumulate.

As a first attempt, I used the global temperature anomaly. The results are shown in Fig. 2. The time constant was rather arbitrarily chosen as 100 years. The best fit was determined by minimizing the rms error between the simulated annual $CO_2$ concentrations, and the observed concentrations, during the period 1959-2006. The other parameters are listed in Table 1: the sensitivity is 0.75 $^oC^{-1}$ for an equilibrium concentration of 435 ppmv at a 0 $^oC$ anomaly. The rms error beween simulated and observed concentrations is only 3.3 ppmv, which is about 1% of the actual concentration.

Two observations emerge from that first attempt. One is that the modeled curve tracks the monthly ΔC curve very well, and both now have a similar trend. On the other hand, it was difficult to get a perfect match with the historical $CO_2$ data. In particular, no parameter set could be found that would match the curvature of the $CO_2$ data. What could explain this is the particular shape of the historical temperature anomaly, which has a significant change in trend during the 1950-1975 period. However, that shape is not the same for all temperature data, but rather, it is mainly a feature of the Northern Hemisphere data. If the oceans are mainly responsible for the behavior that is modeled, the ocean-dominated southern hemisphere data are probably more representative. An even better choice would be the Sea Surface Temperature of the Southern Hemisphere.

Figs. 3 shows the simulation results using the Southern Hemisphere temperature data, and Fig. 4 using the Southern Hemisphere Sea Surface temperature data. Again, the time constant was set at 100 years. Much better agreement was found for those two cases, with an rms error of only 2.0 ppmv in the case of the southern hemisphere data, and 1.1 ppmv for the sea surface temperature. This is almost within the range of experimental measurement errors (+/- 0.4 ppmv). Furthermore, the comparison of monthly data also shows an impressive agreement. The values giving the best fits are similar for all cases, and listed in Table 1. However, other sets of values can give similarly good fits. In particular, if one uses a smaller time constant, the sensitivity, and the equilibrium concentration are also reduced. Too short a time constant, however, leads to larger wiggles in the $CO_2$ concentration curve, that do not match the smoothness of the observations. It therefore appears that the time constant must be at least larger than about 30 years. Fig. 5 shows the simulation results with a 40 year time constant for the sea surface SH temperatures. The rms error is still very low, at 1.3 ppmv. Lower sensitivities also appear more realistic in view of the historical evidence.

The most important, and surprising, finding is that the same set of parameters explains both the long term $CO_2$ concentration trend, and the short term fluctuations in concentration. No compromise was made to maximize the two fits. In fact, the parameters were set by minimizing the rms error on the $CO_2$ concentration curve, and in all cases, this resulted in closely matched monthly anomalies curves. The model, if it represents the actual behavior, single-handedly explains both the long term trend and the short term dynamical behavior.

## Discussion

The first question that comes to mind is certainly: but what about human emissions? There is no mention of them in the model. That is because the system is taken as a whole, and not as a sum of detailed fluxes that individually vary on a monthly, annual, or decadal basis. One does not necessarily need to know where the $CO_2$ comes from to restore the equilibrium. Because of the highly dynamic nature of the carbon cycle, the system takes the $CO_2$ wherever it can to attempt to restore the equilibrium. Since, as has been noted, human emissions are but a small fraction of the total exchanges, they are, in the end, a minor contributor to the cycle, which is easily absorbed by the system. In reality, however, studies of the carbon cycle indicate that the influx of anthopogenic carbon has indeed replaced outgassing as a source of carbon during the industrial age (Raven et al. 1999, Sabine et al. 2004).

A strong argument in favor of human-induced accumulation of $CO_2$ has been the change in isotopic ratios of atmospheric carbon over time. But in the proposed model, since $CO_2$ is accumulating, and human emissions constantly feed the atmosphere with isotopically-modivfied carbon, the isotopic ratio of carbon in the atmosphere is also bound to change. In other words, the accumulation does not need to be caused by human emissions for the ratio to change. It suffices that the human emissions are fed directly to the atmosphere, and not to the other components of the cycle. In a warming world, the oceans are not outgassing $CO_2$ in the atmosphere, they merely fail to uptake the surplus of $CO_2$ that is already there. Of course, the oceans do not stop uptaking carbon at high latitudes and outgassing it at low latitudes. Merely the balance between the two is affected. That almost half the anthropogenic carbon emitted during the anthropocene now finds itself in the bottom of the oceans (Sabine et al. 2004) is only a result of that continuous exchange with the atmosphere, and demonstrates that, in the end, that is where all the carbon ends up, because that is the largest sink.

The "standard" view of the perturbed carbon cycle, adopted for example by the IPCC (Denman 2007), is that of an ultra-stable cycle over the past millenia, suddenly perturbed by human emissions in the $20^{th}$ century. According to that view, the $CO_2$ concentration has remained bounded between 260 and 280 ppm for the last 10,000 years, and atmospheric concentrations have never reached the current level over the past 650,000 years. However, recent reconstructions of past $CO_2$ concentrations, based on leaf stomatal frequencies, point to a different picture (Wegner et al. 2004, Kouwenberg et al. 2005, Jessen et al. 2007). Kouwenberg et al. (2005) have shown that during the past millenium, $CO_2$ concentrations may have fluctuated between as low as 240 ppm and as high as 340 ppm. Moreover, the $CO_2$ concentration also seems to be correlated with the temperature fluctuations during the same period. The same correlation is seen further in the past, for example during the early Holocene (Jessen et al. 2007), where a correlation with solar activity is also found. It is also well known that the $CO_2$ concentration dropped significantly during the glaciations. $CO_2$ also always seems to lag warming, a feature that is also part of the proposed model. Therefore what is proposed here is not a revolutionary way to look at the carbon cycle. It is rather consistent with most of our knowledge about it. The main difference is the role of human emissions. In the standard model, human emissions are an outlandish perturbation, that the system seems unable to manage. But the temporal behavior of the carbon uptake, which seems to be totally independent from the amount of anthropogenic emissions, just cannot be reconciled with that view, whereas it fits well with the proposed model.

Of course, all this begs the question of the actual cause of the 20$^{th}$ century warming, and the role of the $CO_2$ forcing, if any, in that warming. This is not the place to answer those questions, only to speculate that the obvious culprit is the only remaining independent forcing, namely the Sun. Reconstructions of the solar activity show that it has been unusually high during the 20$^{th}$ century (Usoskin et al. 2003, Solanki and al. 2004, Usoskin et al. 2006), and possibly at its highest level since the past 11,000 years. In any case, the climate of the Holocene, for example, seems to be highly correlated with solar activity (Bond et al. 2001). This would call for yet unknown amplifying mechanisms, because the changes of the Total Solar Irradiance (TSI) are too small a forcing to explain the variations in the past climate. The recently proposed link between the cosmic rays and cloud formation could be such a mechanism (Svensmark 2000, Svensmark and Friis-Christensen 1997). The strong effect of the highly variable UV solar irradiance on the stratosphere could also trickle down to trigger important changes in the lower troposphere (Langematz et al. 2005, Labitzke et al. 2006, Bond et al. 2001). One should also not discount the warming effects of various other human activities in the anomalous 20$^{th}$ century warming.

Nevertheless, it is also plausible that the accumulation of $CO_2$ in the atmosphere, and the associated greenhouse effect, act as a feedback on the Sun-induced warming, much in the same way that water vapor is deemed to act. Indeed such a feedback may explain how small changes in solar irradiance trigger deglaciations. In that picture, the total warming would be the combined result of the solar activity, and the ($CO_2$ + water vapor) feedbacks. A feature of the proposed model, however, is that the $CO_2$ concentration takes a long time to reach equilibrium, whereas water vapor comes and goes rapidly. So even if, for example, the solar forcing stops increasing, the $CO_2$ may keep rising for a couple of decades, depending on the time constant, and contribute to a continued warming, which may only be counteracted by the cooling effect of a weaker Sun. An example of this is shown on Fig. 6, where a drop in temperatures at a rate similar to the 20$^{th}$ century increase (0.14 $^o$C/decade) is followed by flat temperatures from 2050 on. The parameters used are those of Fig. 5, in particular the 40 year time constant. The $CO_2$ is seen to rise until 2024, and gradually come down, but would still be at the 320 ppmv level by 2100. This inertia may in fact be the reason why the last decades of the 20$^{th}$ century have seen a continued warming despite a flat to declining solar activity. It may also be the reason why phenomenological models that attribute most of the warming of the first half of the 20$^{th}$ century to solar influences fail to make such an attribution for the second half (Scafetta and West 2006).

## Conclusion

The picture presented here is both compelling and troubling. The idea of a carbon cycle perturbed solely by human emissions is deeply ingrained in our collective thought, and is one of the foundations of the current paradigm. In the alternative, almost heretical view presented here, human emissions are but another perturbation to a system that is entirely driven by its largest sink, as it attempts to reach and maintain equilibrium with the atmosphere. Caution is of course required: this model is still only a hypothesis. The reader has, of course, the liberty of not taking it too seriously. It is, after all, but a "what if" exercise. However, it is not often in science that a very simple model has the kind of fit with two sets of observations as was found here. For that reason alone, the results presented here deserve consideration. In terms of pure empirical support, the proposed model does explain facts and observations that are not accounted for by the standard view.

Future work to prove or disprove this hypothesis may attempt to reconcile the reconstructed $CO_2$ fluctuations of the past centuries with the sensitivity values obtained here. It would also be interesting to see if the evolution of temperatures can be accounted for by a model with increased solar forcing, amplified by the $CO_2$ forcing, with its associated inertia.

## Acknowledgments

Thanks to David Smith for help with the to data sources. Thanks to Arthur Rörsch for sending me his reprints.

## References

Bousquet, P., Peylin, P. Ciais, P. Le Quéré, C. Friedlingstein, P. Tans, P. P., 2000, Regional Changes in Carbon Dioxide Fluxes of Land and Oceans Since 1980, Science, vol. 290, pp. 1342-1346.

Denman, K.L., G. Brasseur, A. Chidthaisong, P. Ciais, P.M. Cox, R.E. Dickinson, D. Hauglustaine, C. Heinze, E. Holland, D. Jacob, U., Lohmann, S Ramachandran, P.L. da Silva Dias, S.C. Wofsy and X. Zhang, 2007: Couplings Between Changes in the Climate System and Biogeochemistry. In: *Climate Change 2007: The Physical Science Basis. Contribution of Working Group I to the Fourth Assessment Report of the Intergovernmental Panel on Climate Change* [Solomon, S., D. Qin, M. Manning, Z. Chen, M. Marquis, K.B. Averyt, M.Tignor and H.L. Miller (eds.)]. Cambridge University Press, Cambridge, United Kingdom and New York, NY, USA.

Jessen, C.A., Rundgren, M., Björk, S., Muscheler, R., 2007, Climate forced atmospheric $CO_2$ variability in the early Holocene: A stomatal frequency reconstruction, Global and Planetary Change, vol. 57, pp. 247-260.

Kouwenberg, L. Wagner, R., Kürschner, W. Visscher, H., 2005, Atmospheric $CO_2$ fluctuations during the last millennium reconstructed by stomatal frequency analysis of Tsuga heterophylla needles, Geology, vol. 33, pp. 33-36.

Labitzke K., and van Loon, H., Solar effects in the middle and lower stratosphere and probable associations with the troposphere, In: Space Weather - Physics and Effects, (Eds.: Volker Bothmer and Ioannis A. Daglis), Chapter 8, 225 - 245, Springer Verlag, 2006.

Langematz, U., Matthes, K., Grenfell, J. L., 2005, Solar impact on the climate : modeling the coupling between the middle and the lower atmosphere, Mem. S. A. It. Vol. 76, pp. 868-875.

MacRae, A.M.R., 2008 : Carbon dioxide is not the primary cause of global warming: the future can not cause the past, http://icecap.us/images/uploads/CO2vsTMacRae.pdf

Raven, J. A., Falkowski, P.G., 1999, Oceanic sinks for atmospheric $CO_2$, Plant, Cell and Environment vol. 22, pp. 741-755.

Rörsch A., Courtney R. S. and Thoenes D., 2005, The interaction of climate change and the carbon dioxide cycle, Energy&Environment 16(2), 2005, 217-238.

Sabine, C.L., Feely, R.A., Gruber, N., Key, R. M., Lee, K. Bullister, J.L., Wanninkhof, R., Wong, C. S., Wallace, D. W. R., Tlibrook, B., Millero, F. J., Peng, T., Kozyr, A., Ono, T., Rios, A. F., 2004, The Oceanic Sink for Anthropogenic $CO_2$, Science vol. 305, pp. 367-371.

Sarmiento, J. L., Hughes, T. M. C., Stouffer, R. J. Manabe, S., 1998, Simulated response of the ocean carbon cycle to anthropogenic climate warming, Nature, vol. 393, pp. 245-248.

Scafetta, N., West, B. J., 2006, Phenomenological solar contribution to the 1900-2000 global surface warming, Geophysical Research Letters, vol. 33, L05708.

Solanki, S. K., Usoskin, I. G., Kromer, B. Schüssler, M. Beer, J., 2004, Unusual activity of the Sun during recent decades compared to the previous 11,000 years, Nature, vol. 431, pp. 1084-1086.

Svensmark, H., Friis-Christensen, E., 1997, Variation in cosmic rays flux and global cloud coverage: a missing link in solar climate relationship, Journal of Atmospheric and Solar-Terrestrial Physics, vol. 59, pp. 1225-1232.

Svensmark, H., 2000, Cosmic rays and Earth's climate, Space Science Reviews vol. 93, pp. 155-166.

Usoskin, I.G., 2003, Millennium-Scale Sunspot Number Reconstruction: Evidence for an Unusually Active Sun since the 1940's, Physical Review Letters, vol. 91, 211101.

Usoskin, I. G., Solanki, S. K., Korte, M., 2006, Solar activity reconstructed over the last 7000 years : the influence of geomagnetic field changes, Geophysical Research Letters, vol. 33, L08103.

Wagner, F., Kouwenberg, L. L. R., van Hoof, T. B., Visscher, H., 2004, Reprocucibility of Holocene atmospheric $CO_2$ records based on stomatal frequency, Quart. Science Rev. 23, 1947-1954.

| Parameter | Global | SH | SST-SH | SST-SH |
|---|---|---|---|---|
| $C_{e0}$ (ppmv) | 435 | 448 | 471 | 391 |
| $\gamma$ (°C$^{-1}$) | .75 | .79 | .77 | .555 |
| $\alpha$ (years) | 100 | 100 | 100 | 40 |
| C in 1885 (ppmv) | 280 | 280 | 280 | 280 |
| rms error (ppmv) for annual $CO_2$ 1959-2006 | 3.3 | 2.0 | 1.1 | 1.3 |

Table 1 : Parameters used for the modeling in figs. 2-5

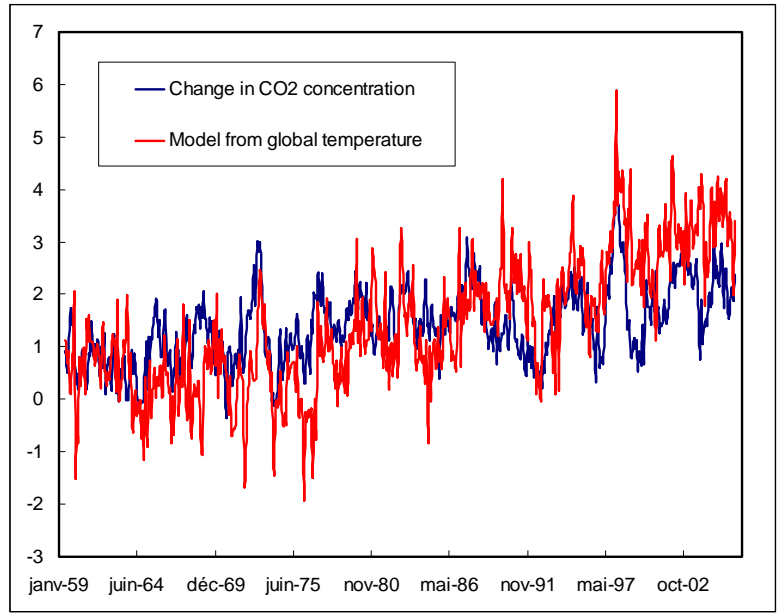

**Fig. 1** Comparison of the monthly annualized change in $CO_2$ concentration with modeled changes, using a linear temperature dependence model.

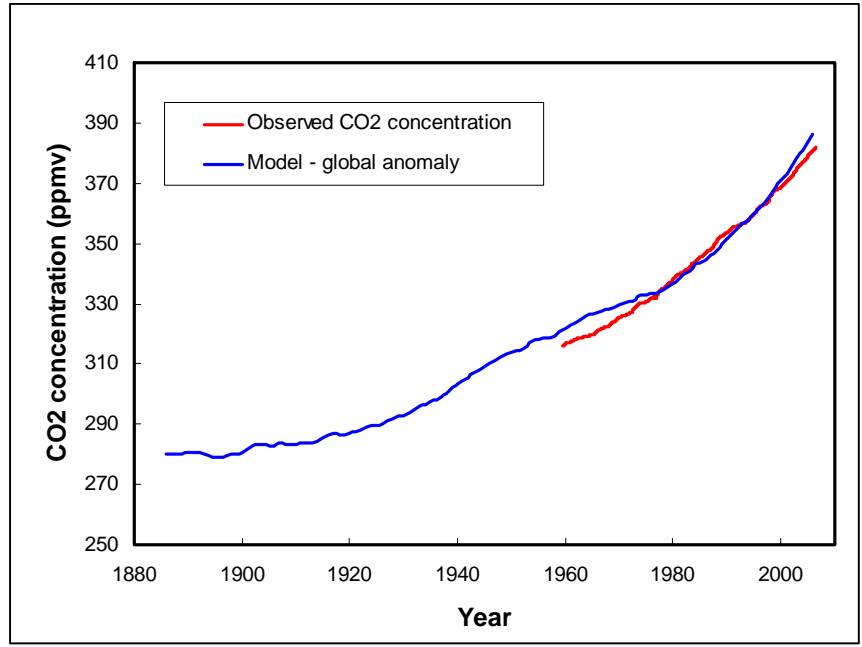

(a)

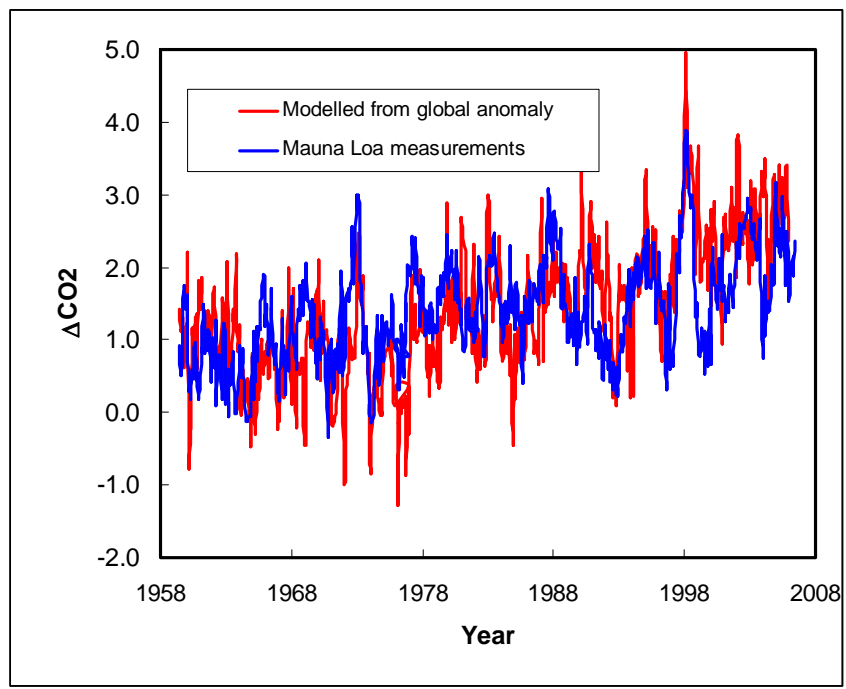

(b)

Fig. 2 (a) Comparison of the historical $CO_2$ concentrations with those modeled from the global temperature anomaly; (b) Comparison of the monthly annualized changes in atmospheric $CO_2$ concentration with those modeled using the global temperature anomaly.

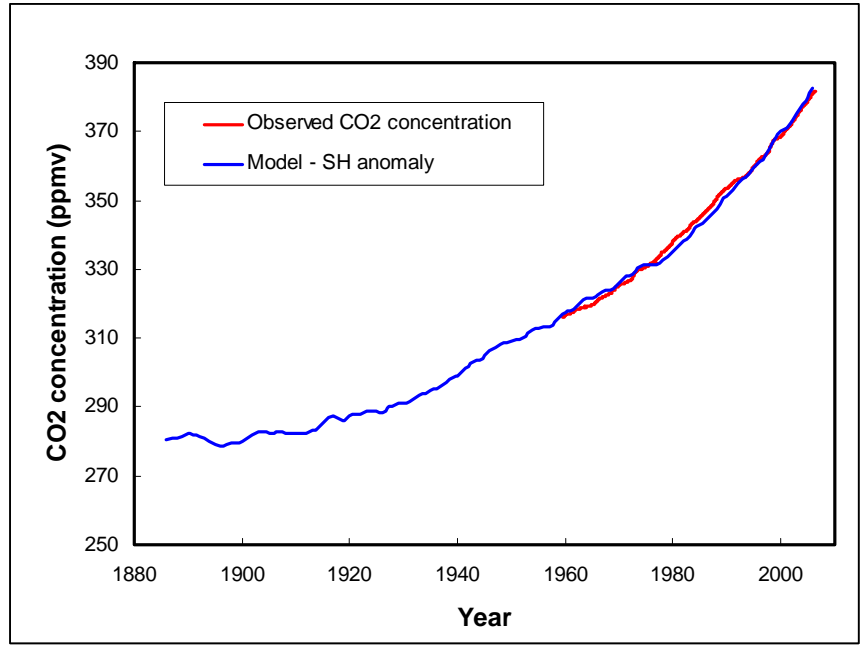

(a)

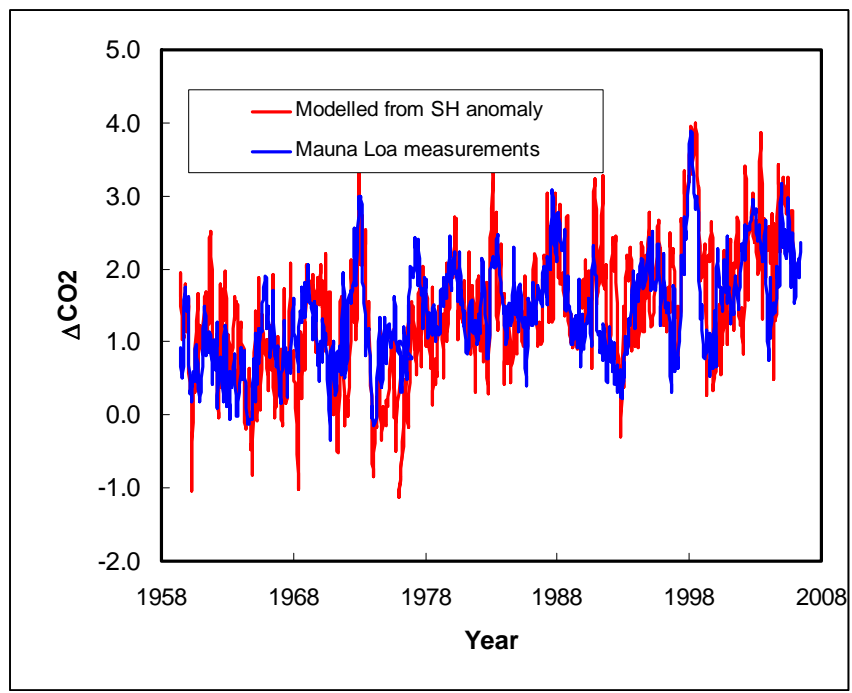

(b)

**Fig. 3** (a) Comparison of the historical $CO_2$ concentrations with those modeled from the southern hemisphere temperature anomaly; (b) Comparison of the monthly annualized changes in atmospheric $CO_2$ concentration with those modeled using the southern hemisphere temperature anomaly.

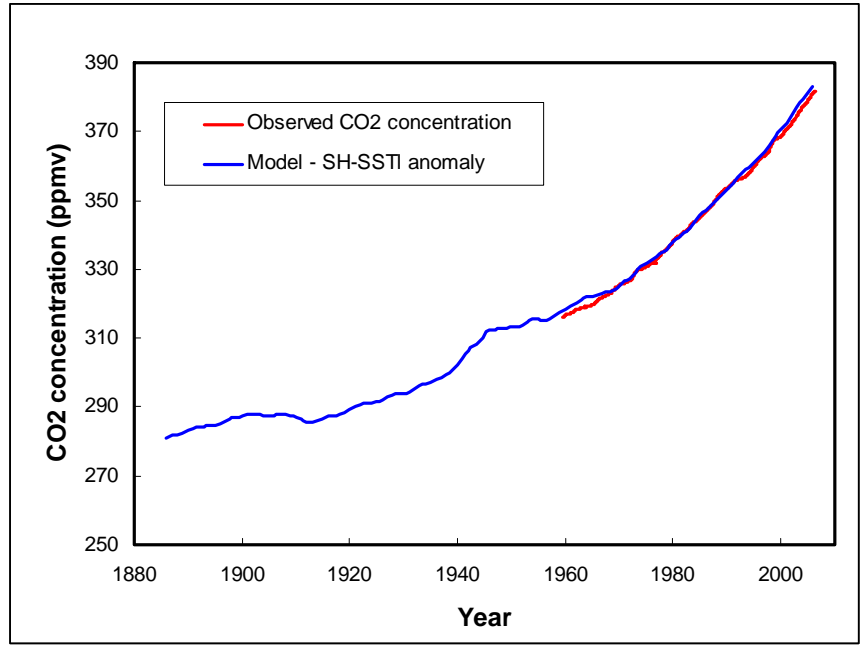

(a)

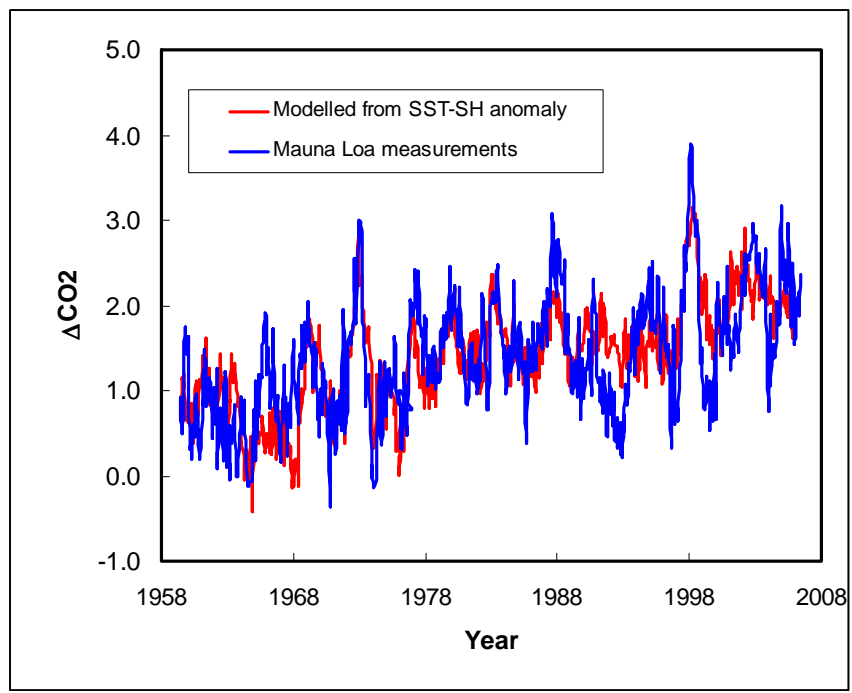

(b)

Fig. 4  (a) Comparison of the historical $CO_2$ concentrations with those modeled from the SST-SH temperature anomaly; (b) Comparison of the monthly annualized changes in atmospheric $CO_2$ concentration with those modeled using the SST-SH temperature anomaly. The time constant is 100 years.

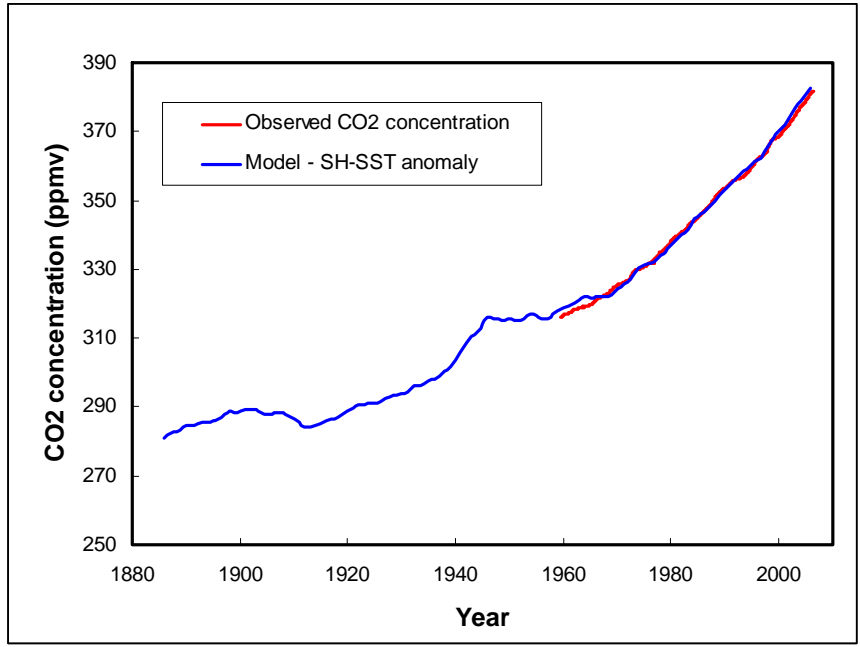

(a)

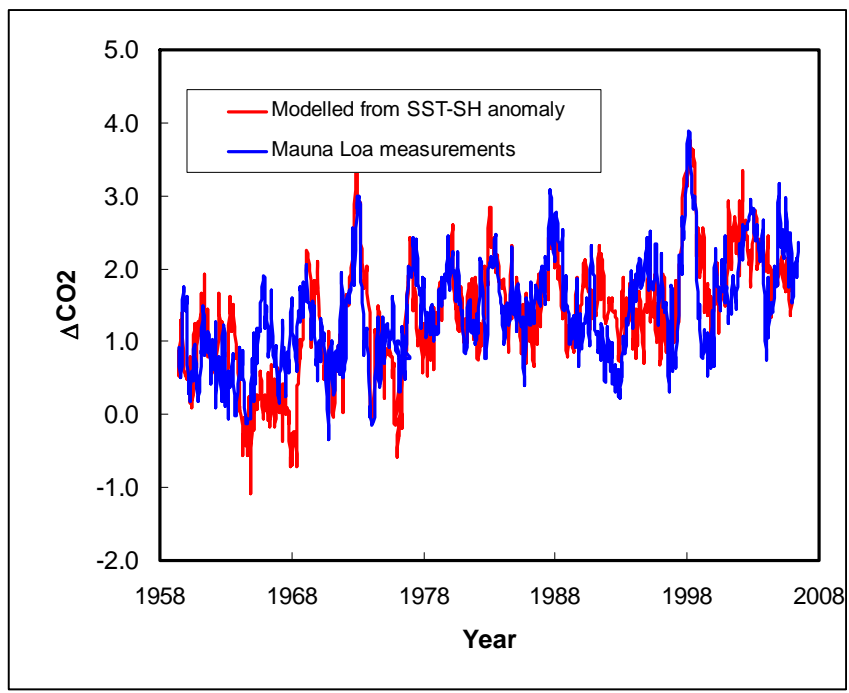

(b)

Fig. 5 (a) Comparison of the historical $CO_2$ concentrations with those modeled from the SST-SH temperature anomaly; (b) Comparison of the monthly annualized changes in atmospheric $CO_2$ concentration with those modeled using the SST-SH temperature anomaly. The time constant is 40 years.

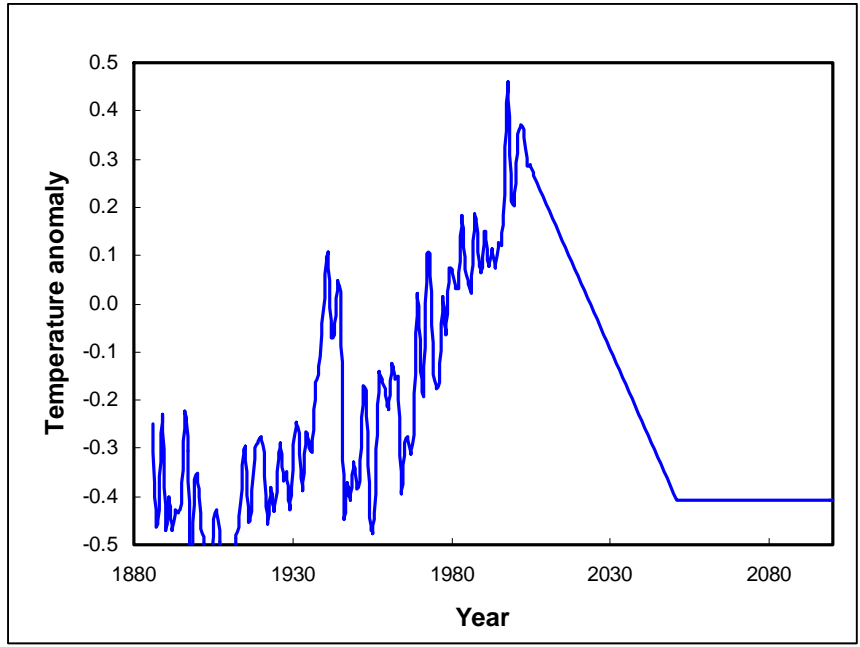

(a)

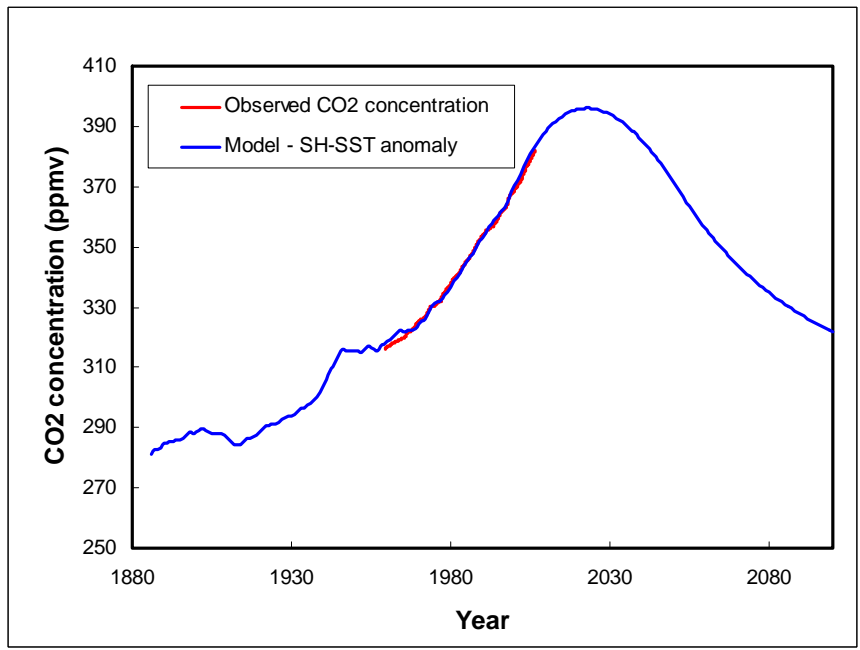

(b)

Fig. 6  Hypothetical evolution of $CO_2$ concentration following a drop in temperatures, using the parameters of Fig. 5. (a) temperature anomaly; (b) $CO_2$ concentration.